\documentclass[seceq]{ptptex}
\usepackage{wrapft}
\usepackage{bm}

\title{
Topological Grand Unification
}

\author{
Yoshiharu \textsc{Kawamura}\footnote{E-mail: haru@azusa.shinshu-u.ac.jp}
}

\inst{
Department of Physics, Shinshu University, Matsumoto 390-8621, Japan
}

\recdate{
October 15, 2008}

\abst{
We propose a new grand unification scenario for ensuring 
proton stability and triplet-doublet Higgs mass splitting 
with the help of topological symmetry and dynamics.
}

\begin{document}

\maketitle

\section{Introduction}

Grand unification is an attractive idea and enables the unification of forces 
and the (partial) unification of quarks and leptons in each family.\cite{GUT}
For gauge bosons and the weak Higgs doublet in the standard model (SM),
extra gauge bosons and colored Higgs bosons are necessary
to form complete multiplets of a unified gauge group, $G_U$.
Extra particles and their superpartners cause severe problems as if they have come from Pandora's box. 
The typical ones are the proton decay problem (Why is the proton so stable?)\cite{Proton}
and the triplet-doublet Higgs mass splitting problem 
(How is the Higgs mass splitting realized without fine-tuning among parameters?),\footnote{
There have been several interesting proposals for solving the triplet-doublet Higgs mass splitting problem 
by extending the model.\cite{sliding}$^-$\cite{discrete}}
in the minimal supersymmetric extension of the grand unified theory (SUSY GUT).\cite{SUSYGUT}

The root of all evils stems from the existence of extra heavy particles that generate physical effects,
and hence we come across the conjecture that
{\it the proton stability and the triplet-doublet Higgs mass splitting can be realized
to be compatible with the grand unification,
if extra particles are unphysical and affect no physical processes.}
In other words, 
gauge coupling unification originates from a unified gauge symmetry on a high-energy scale
and fields are $G_U$-multiplets.
If some of them had unphysical degrees of freedom, 
no full symmetry would be realized in the physical world.
Then physical fields can be multiples of the SM gauge group $G_{\tiny{\rm SM}}$
and can survive on the low-energy scale, where SM particles (and their superpartners) play an essential role.
Unphysical modes would be eliminated by a large local symmetry 
such as topological symmetry \cite{TFT}and dynamics.

In this paper, we propose a new grand unification scenario for ensuring
proton stability and triplet-doublet Higgs mass splitting
on the basis of the above conjecture. 
We give simple models with an $SU(5)$ gauge group on a space-time 
including extra superspace to realize our proposal.
The reduction to the SM or minimal supersymmetric SM (the MSSM) is carried out 
using the Parisi-Sourlas mechanism.\cite{PS}

This paper is organized as follows.
In the next section, we elaborate our scenario.
In \S 3, we construct simple models to materialize our scenario.
In \S 4, we present conclusions and discussion.

\section{Our scenario}

SM particles consist of three kinds of gauge bosons $(G_{\mu}, W_{\mu}$, and $B_{\mu})$,
three families of quarks and leptons, and the weak Higgs doublet $h_W$.
The minimal unification of gauge bosons is realized using the $SU(5)$ gauge group as follows.
By introducing extra gauge bosons ($X_{\mu}, Y_{\mu}$, $\bar{X}_{\mu}$, and $\bar{Y}_{\mu}$) 
and the colored Higgs boson $h_C$,
complete $SU(5)$ multiplets are formed as
\begin{eqnarray}
&~& A^{\alpha}_{\mu} = G_{\mu} + W_{\mu} + B_{\mu} 
 + (X_{\mu}, Y_{\mu}) + (\bar{X}_{\mu}, \bar{Y}_{\mu}) ~;
\nonumber \\
&~&  ~~ {\bf 24} = ({\bf 8}, {\bf 1})_0 + ({\bf 1}, {\bf 3})_0 + ({\bf 1}, {\bf 1})_0   
 + ({\bf 3}, {\bf 2})_{-5/6} + (\bar{\bf 3}, {\bf 2})_{5/6} ,
\label{24}\\
&~& h = h_C + h_W ~; ~~ {\bf 5} = ({\bf 3}, {\bf 1})_{-1/3} + ({\bf 1}, {\bf 2})_{1/2}, 
\label{5}
\end{eqnarray}
where the boldface numbers represent gauge quantum numbers under $SU(5)$ and 
$G_{\tiny{\rm{SM}}} = SU(3)_C \times SU(2)_L \times U(1)_Y$.
On the other hand, quarks and leptons are organized as three kinds of multiplets, namely, 
${\bf 1}$, $\bar{\bf 5}$ and ${\bf 10}$ for each family,
\begin{eqnarray}
&~& \psi_{{\bf 1}L} = (\nu_R)^c ~; ~~ {\bf 1} = ({\bf 1}, {\bf 1})_0 ,
\label{1}\\
&~& \psi_{{\bar{\bf 5}}L} = (d_R)^c + l_L ~; ~~ \bar{\bf 5} = (\bar{\bf 3}, {\bf 1})_{1/3} + ({\bf 1}, {\bf 2})_{-1/2} , 
\label{bar5}\\
&~& \psi_{{\bf 10}L} = (u_R)^c + (e_R)^c + q_L ~; ~~ 
{\bf 10} = (\bar{\bf 3}, {\bf 1})_{-2/3} + ({\bf 1}, {\bf 1})_1 + ({\bf 3}, {\bf 2})_{1/6} ,
\label{10}
\end{eqnarray}
where the family index is omitted for simplicity.

Our standpoint is that the grand unification is realized based on a simple gauge group, $G_U$.
Then the gauge coupling unification comes from the fact that
the gauge coupling $g_U$ is unique in $G_U$, and 
various fields are unified by forming $G_U$-multiplets.
Our starting point is an action integral with a manifest unified gauge symmetry, including $G_U$-multiplets.
The appearance of extra particles can cause serious problems
such as the proton decay problem and the triplet-doublet Higgs mass splitting problem. 
Conventionally, these problems are solved through the extension of the model,
leaving dangerous particles physical.

We explore a new possibility of constructing a problem-free model.
We assume that extra particles are unphysical 
and have no effect on any physical processes even at the quantum level.
Then no full symmetry is realized in the physical world.
Physical fields are $G_{\tiny{\rm{SM}}}$-multiples 
and survive in low-energy physics, where SM particles (and their superpartners) play an essential role.
In order to induce such a reduction, a large local symmetry is absolutely imperative.
A possible candidate is a topological symmetry.\cite{TFT}
Topological symmetry is realized by a conserved charge with nilpotency
and is interpreted as BRST symmetry.
Under the transformation, bosonic (Grassmann even) objects are transformed into 
fermionic (Grassmann odd) objects with identical spins.
We refer to topological symmetry as BRST symmetry or supersymmetry in this paper, 
although it might be confused with the space-time supersymmetry.

We assume that a Lagrangian density $\mathcal{L}_U$, which possesses a BRST symmetry
and a unified gauge symmetry, is dynamically derived from a more fundamental theory.
Our goal is that we arrive at the SM or MSSM after eliminating unphysical modes.
Specifically, after eliminating extra degrees of freedom including extra coordinates,
we would like to obtain the action integral of the SM or MSSM 
\begin{eqnarray}
\int d^n X {\mathcal{L}}_{U} \Rightarrow \int d^4x {\mathcal{L}}_{\tiny{\rm SM}} ~~
{\rm or} ~~ \int d^4x {\mathcal{L}}_{\tiny{\rm MSSM}} ,
\label{LSM}
\end{eqnarray}
where $X$ represents an extented space-time coordinate and ${\mathcal{L}}_{\tiny{\rm SM}}$ 
$({\mathcal{L}}_{\tiny{\rm MSSM}})$ is the SM (the MSSM) Lagrangian density.
If this program comes off, proton stability and triplet-doublet Higgs mass splitting 
can be realized to be compatible with grand unification.
The unification scale $M_U$ is regarded as an energy scale on which 
the unified theory described by ${\mathcal{L}}_U$ is derived dynamically
or the reduction to the SM (or MSSM) occurs.
We refer to our scenario as {\it topological grand unification} or {\it unphysical grand unification}.

Topological grand unification can be regarded as the most extreme generalization of grand unification on an orbifold,
where unwelcome massless modes are projected out by orbifolding
and triplet-doublet Higgs mass splitting is realized.\cite{orb}
However, Kaluza-Klein modes survive after dimensional reduction and they could be seeds of trouble.
In our scenario, we expect that such particles become unphysical and harmless by a large local symmetry
and a dynamics fixing a unified theory on an extra space.

\section{Models}

We discuss models with the $SU(5)$ gauge group on a space-time 
including extra commuting and anticommuting dimensions to realize our proposal.
Models possess supersymmetry, which eliminates unphysical degrees of freedom.
If fields assume suitable configurations on an extra space,
the SM or MSSM is expected to appear through dimensional reduction.

\subsection{Superspace}

Space-time is assumed to be factorized into a product of 4D Minkowski space-time $(M^4)$
and superspace $(\Omega)$ including 2D Euclidean space $(R^2)$ and 2D fermionic space,
whose coordinates are denoted by $x^{\mu}$ $(\mu = 0, 1, 2, 3)$ 
and $(y^j, \theta, \bar{\theta})$ $(j = 1, 2)$, respectively.
The notation $X^M = (x^{\mu}, y^j, \theta, \bar{\theta})$ is also used.
$\theta$ and $\bar{\theta}$ are Grassmann numbers satisfying 
\begin{eqnarray}
\theta^{\dagger} = -\theta , ~~ \bar{\theta}^{\dagger} = - \bar{\theta} , ~~
\theta^2 = \bar{\theta}^2 =0 , ~~ \theta \bar{\theta} = - \bar{\theta} \theta , ~~
\int d\theta \theta = \int d\bar{\theta} \bar{\theta} = i .
\label{theta}
\end{eqnarray}
The inner product of the two vectors $X^M_{(1)}$ and $X^M_{(2)}$ is defined by
\begin{eqnarray}
 \eta_{MN} X^M_{(1)} X^N_{(2)} \equiv \eta_{\mu\nu} x^{\mu}_{(1)} x^{\nu}_{(2)} 
 + y^1_{(1)} y^1_{(2)} + y^2_{(1)} y^2_{(2)}
 + \frac{2}{\mu^2} \left(\bar{\theta}_{(1)} \theta_{(2)} + \bar{\theta}_{(2)} \theta_{(1)}\right) ,
\label{innerproduct}
\end{eqnarray}
where $\mu^2$ is a parameter and the nonvanishing components of metric concerning Grassmann coordinates are given by
\begin{eqnarray}
\eta_{\theta \bar{\theta}} = - \frac{2}{\mu^2} , ~~ \eta_{\bar{\theta} \theta} = \frac{2}{\mu^2} .
\label{eta}
\end{eqnarray}
The inner product (\ref{innerproduct}) is invariant under the following transformation:
\begin{eqnarray}
&~& y^j_{(r)} \to y'^j_{(r)}  = y^j_{(r)}  + 2 \bar{a}^j \xi \theta_{(r)} + 2 {a}^j \xi \bar{\theta}_{(r)}   , 
\nonumber\\
&~& \theta_{(r)} \to \theta'_{(r)} = \theta_{(r)} + \mu^2 a^j y_{j(r)} \xi , ~~
\bar{\theta}_{(r)} \to \bar{\theta}'_{(r)} = \bar{\theta}_{(r)} - \mu^2 \bar{a}^j y_{j(r)} \xi ,
\label{BRST-SUSY}
\end{eqnarray}
where $r=1,2$, $a^j$ and $\bar{a}^j$ are arbitrary parameters and $\xi$ is a Grassmann number.
Transformations form the orthosymplectic supergroup $OSp(2/2)$ including the 2D orthogonal group $O(2)$
and symplectic group $OSp(2)$ as a subgroup.

The superfield $\Phi(x, y, \theta, \bar{\theta})$ is a field on $M^4 \times \Omega$ and is expanded as
\begin{eqnarray}
\Phi(x, y, \theta, \bar{\theta}) = \phi(x, y) + \theta \chi(x, y) 
 + \bar{\theta} \eta(x, y)  + \bar{\theta}\theta \varphi(x, y) .
\label{BRST-SF}
\end{eqnarray}
When the operator $\Psi(x, y, \theta, \bar{\theta})$ takes the
manifest $OSp(2/2)$ invariant form,
\begin{eqnarray}
\Psi(x, y, \theta, \bar{\theta}) = \psi(x, y^2 +(4/\mu^2)\bar{\theta}\theta) 
= \psi(x, y^2) 
+ \frac{4}{\mu^2} \bar{\theta}\theta \frac{\partial}{\partial y^2} \psi(x, y^2) ,
\label{BRST-SUSY-inv}
\end{eqnarray}
the integral of $\Psi(x, y, \theta, \bar{\theta})$ over $M^4 \times \Omega$ turns out to be the
4D integral over $M^4$ as
\begin{eqnarray}
&~& I_{8D} \equiv \int d^4x \int d^2y \int d\theta d\bar{\theta} \Psi(x, y, \theta, \bar{\theta})
\nonumber \\
&~& ~~~~~ = \int d^4x \int d^2y \int d\theta d\bar{\theta} \left(\psi(x,y^2)
 + \frac{4}{\mu^2} \bar{\theta}\theta \frac{\partial}{\partial y^2} \psi(x, y^2)\right)
\nonumber \\
&~& ~~~~~  =  - \frac{4}{\mu^2} \int d^4x \int d^2y \frac{\partial}{\partial y^2} \psi(x, y^2)  
 = - \frac{4\pi}{\mu^2} \int d^4x \int_0^{\infty} dr^2 \frac{\partial}{\partial r^2} \psi(x, r^2) 
\nonumber \\
&~& ~~~~~  =  \frac{4\pi}{\mu^2} \int d^4x \psi(x, 0) \equiv I_{4D} ,
\label{4D-S}
\end{eqnarray}
where we assume that $\psi(x, r^2)$ vanishes at infinity ($r \to \infty$) on $\Omega$.
This kind of dimensional reduction mechanism is called the Parisi-Sourlas mechanism.\cite{PS}

\subsection{Emergent SM}

Now let us construct a model with $SU(5)$ gauge symmetry on the space-time $M^4 \times \Omega$
and derive the SM through dimensional reduction, under ansatz for field configurations.
Our main players are the $SU(5)$ gauge field $A_M(x,y,\theta,\bar{\theta}) = A_M^{\alpha}(X) T^{\alpha}$, 
the Higgs field ~$h(x,y,\theta,\bar{\theta})$~ and the ~$n_g$~ families of matter fields 
$(\psi_{{\bf 1}}(x,y,\theta,\bar{\theta})$, 
$\psi_{\bar{\bf 5}}(x,y,\theta,\bar{\theta})$, $\psi_{{\bf 10}}(x,y,\theta,\bar{\theta}))$ 
whose representations are ${\bf 24}$, ${\bf 5}$ and (${\bf 1}$, $\bar{\bf 5}$, ${\bf 10}$), respectively.
Here, $T^{\alpha}$s are $SU(5)$ gauge generators classified into two sets
(the SM gauge generators $T^a$ and the other gauge generators $T^{\hat{a}}$):
\begin{eqnarray}
&~& T^a ~~ (a = 1, \cdots, 8, 21, \cdots, 24) ~~
\Leftrightarrow ~~ ({\bf 8}, {\bf 1})_0 + ({\bf 1}, {\bf 3})_0 + ({\bf 1}, {\bf 1})_0  , 
\nonumber \\
&~& T^{\hat{a}} ~~ (\hat{a} = 9, \cdots, 20) ~~
\Leftrightarrow ~~  ({\bf 3}, {\bf 2})_{-5/6} + (\bar{\bf 3}, {\bf 2})_{5/6} .
\label{Ta}
\end{eqnarray}
The matter fields contain 6D Weyl spinors corresponding to Dirac spinors on $M^4$,
i.e., $\psi_{{\bm n}} = (\psi_{{\bm n}L}, \psi_{{\bar{\bm n}}L}) = (\psi_{{\bm n}L}, \psi_{{\bm n}R})$
$({\bm n} = {\bf 1}$, $\bar{\bf 5}$, ${\bf 10})$.\footnote{
Gauge theories including scalar fields and spinor fields
on superspace of 6 commuting and 2 anticomuting dimensions were formulated in \citen{MNTW}.}

A general $OSp(2/2)$ supersymmetric and $SU(5)$ gauge invariant action integral on $M^4 \times \Omega$ is given by
\begin{eqnarray}
&~& S_{SU(5)} = \int d^4x \int d^2y \int d\theta d\bar{\theta} L_{SU(5)}(x, y, \theta, \bar{\theta}) ,
\nonumber \\
&~&   L_{SU(5)}(x, y, \theta, \bar{\theta}) =  -\frac{1}{2}{\rm tr}F^{MN}F_{MN} + (D^M h)^{\dagger} (D_M h) -V(h)
\nonumber \\
&~& ~~~~~~~~~~~~ + \sum_{n_g~{\rm families}} \left(i \bar{\psi}_{\bar{\bf 5}} \Gamma^M D_M \psi_{\bar{\bf 5}}
+ i \bar{\psi}_{{\bf 10}} \Gamma^M D_M \psi_{{\bf 10}}\right. 
\nonumber \\
&~& ~~~~~~~~~~~~~~~~~~~~~~~~~~ \left. + f_{U} h {\psi}_{{\bf 10}} {\psi}_{{\bf 10}} 
 + f_{D} h^{*} {\psi}_{{\bf 10}} \psi_{\bar{\bf 5}} + {\rm h.c.}\right) ,
\label{S-SU5}
\end{eqnarray}
where $F_{MN}$ is the superspace field strength of $A_M$, $D_M = \partial_M + ig_U A_M$,
$V(h)$ is the Higgs potential, $\Gamma^M$ is the superspace gamma matrices,
$f_U$ and $f_D$ are Yukawa coupling matrices,
and h.c. is the hermitian conjugate of the former terms.
Here and hereafter we omit the gauge singlet neutrinos for simplicity.

Let us assume that field configurations on $\Omega$ are dynamically fixed (up to a freedom of 
residual gauge symmetry) on the scale $M_U$ in a more fundamental theory.
To realize our scenario, we assume that fields take the following configurations:
\begin{eqnarray}
&~& A^{\mu}(X) = A^{\mu}(x, w^2) , ~~
A^{y^j}(X) = y^j a(x, w^2) , ~~
\nonumber \\
&~& A^{\theta}(X) = {\theta} a(x, w^2) , ~~
A^{\bar{\theta}}(X) =  \bar{\theta} a(x, w^2) , ~~
\label{A-ansatz}\\
&~& h(X) = h(x, w^2) , ~~ \psi_{{\bm n}}(X) = \psi_{{\bm n}}(x,w^2) ,
\label{H-ansatz}
\end{eqnarray}
where $w^2 \equiv y^2 +(4/\mu^2)\bar{\theta}\theta$.
At this stage, $\psi_{{\bm n}}$s do not possess the transformation property of $OSp(2/2)$ spinors.
Furthermore, every field lacks the translational invariance on $\Omega$.
The gauge-transformed configurations are equivalent to the original ones
with a restricted gauge transformation function, $U(x,w^2)=\exp(i\xi^{\alpha}(x,w^2)T^{\alpha})$.
The gauge transformations for gauge bosons and Higgs boson are given by
\begin{eqnarray}
&~& A_{\mu}(x,w^2) \to A'_{\mu}(x,w^2) 
= U(x,w^2) A_{\mu}(x,w^2) U^{-1}(x,w^2) 
\nonumber \\
&~& ~~~~~~~~~~~~~~~~~~~~~~~~~~~~~~~~~~~~ - \frac{i}{g_U} U(x,w^2) \partial_{\mu} U^{-1}(x,w^2) ,
\nonumber \\
&~& h(x,w^2) \to h'(x,w^2) = U(x,w^2) h(x,w^2) .
\label{gauge-inv}
\end{eqnarray}
The transformations for matter fields are given in the same way.

Under the above ansatz (\ref{A-ansatz}) and (\ref{H-ansatz}), 
the action integral (\ref{S-SU5}) is rewritten as 
\begin{eqnarray}
&~& S_{SU(5)} = \int d^4x \int d^2y \int d\theta d\bar{\theta} {\mathcal{L}}(x, w^2) ,
\nonumber \\
&~&   {\mathcal{L}}(x, w^2)  =  -\frac{1}{2}{\rm tr}F^{\mu\nu}(x,w^2)F_{\mu\nu}(x,w^2)
\nonumber \\
&~& ~~~~~~~ + w^2\left(\partial^{\mu} a - 2 \frac{\partial}{\partial w^2}A^{\mu}
 + [A^{\mu}, a]\right)\left(\partial_{\mu} a - 2 \frac{\partial}{\partial w^2}A_{\mu} + [A_{\mu}, a]\right)
\nonumber \\
&~& ~~~~~~~ + (D^{\mu} h(x,w^2))^{\dagger} (D_{\mu} h(x,w^2)) - V(h(x,w^2))
 + w^2\left(1 + g^2 a(x, w^2)^2\right) |h(x,w^2)|^2
\nonumber \\
&~& ~~~~~~~ + \sum_{n_g~{\rm families}} 
 \left(i \bar{\psi}_{\bar{\bf 5}}(x,w^2) \Gamma^{\mu} D_{\mu} \psi_{\bar{\bf 5}}(x,w^2)
 + i \bar{\psi}_{{\bf 10}}(x,w^2) \Gamma^{\mu} D_{\mu} \psi_{{\bf 10}}(x,w^2)\right. 
\nonumber \\
&~& \left.  + f_{U} h(x,w^2) {\psi}_{{\bf 10}}(x,w^2) {\psi}_{{\bf 10}}(x,w^2) 
+ f_{D} h^{*}(x,w^2) {\psi}_{{\bf 10}}(x,w^2) \psi_{\bar{\bf 5}}(x,w^2) + {\rm h.c.}\right) .
\label{S-8D}
\end{eqnarray}
The above action integral (\ref{S-8D}) is still $OSp(2/2)$-supersymmetric.
Our assumtion is rephased such that {\it an action integral with specific field configurations on $\Omega$  
such as (\ref{S-8D}) is dynamically derived on the scale $M_U$ from a more fundamental theory.}\footnote{
Conventionally, field configurations are determined by solving field equations 
under appropriate boundary conditions.
It has not been clarified how the configurations (\ref{A-ansatz}) and (\ref{H-ansatz}) are derived.}
After the Parisi-Sourlas dimensional reduction,
we derive the 4D action integral
\begin{eqnarray}
&~& S_{SU(5)} = \int d^4x {\mathcal{L}}(x, 0) \equiv S_{4D} ,
\nonumber \\
&~&   {\mathcal{L}}(x, 0)  =  -\frac{1}{2}{\rm tr}F^{\mu\nu}(x,0)F_{\mu\nu}(x,0) 
+ (D^{\mu} h(x,0))^{\dagger} (D_{\mu} h(x,0)) -V(h(x,0))
\nonumber \\
&~& ~~~~ + \sum_{n_g~{\rm families}} \left(i \bar{\psi}_{\bar{\bf 5}}(x,0) \Gamma^{\mu} D_{\mu} \psi_{\bar{\bf 5}}(x,0)
+ i \bar{\psi}_{{\bf 10}}(x,0) \Gamma^{\mu} D_{\mu} \psi_{{\bf 10}}(x,0)\right. 
\nonumber \\
&~& ~~~~ \left. + f_{U} h(x,0) {\psi}_{{\bf 10}}(x,0) {\psi}_{{\bf 10}}(x,0) 
 + f_{D} h^{*}(x,0) {\psi}_{{\bf 10}}(x,0) \psi_{\bar{\bf 5}}(x,0) + {\rm h.c.}\right) ,
\label{S-4D}
\end{eqnarray}
where we assume that ${\mathcal{L}}(x, r^2)$ vanishes at infinity ($r \to \infty$) on $\Omega$
and we take $\mu^2 = 4\pi$.
As shown by (\ref{S-4D}), the field configurations around the origin on $\Omega$ determine the theory on $M^4$.
Specifically, if only SM fields survive at the origin on $\Omega$,
\begin{eqnarray}
&~& A^{a}_{\mu}(x,0) \ne 0 , ~~ h_W(x,0) \ne 0 , ~~
(d_R)^c_i(x,0) \ne 0, ~~ l_{Li}(x,0) \ne 0 ,
\nonumber \\
&~& (u_R)^c_i(x,0) \ne 0, ~~  (e_R)^c_i(x,0) \ne 0, ~~ q_{Li}(x,0) \ne 0,  
\nonumber \\
&~& A^{\hat{a}}_{\mu}(x,0) = 0 , ~~ h_C(x,0) = 0 , \cdots ,
\label{conf}
\end{eqnarray}
the action integral $S_{4D}$ turns out to be the SM one without the QCD $\theta$ term.
The absence of the strong {\it CP}-violating term is due to the fact that there is no term
in the action integral (\ref{S-SU5}) to generate it.\footnote{
This reasoning is the same as that in Ref.~\citen{2T}.
In Ref.~\citen{2T}, the emergency of SM (or its dual theory) from a higher-dimensional theory 
has been discussed in the framework of 2T-physics.
Their idea is similar to ours, but the starting point and guiding principle are different from ours.}
Hence, the strong {\it CP} problem can be solved if the value of ${\rm argdet}(M_u M_d)$ is sufficiently suppressed
where $M_u$ and $M_d$ are mass matrices of the up- and down-type quarks, respectively.
In (\ref{conf}), $i$ is the family index $(i = 1, 2, 3)$
and the ellipsis represents the disappearance of other fermions at $w^2 = 0$.
We assume that a subset of each $SU(5)$ multiple, in general, survives at $w^2 = 0$ and 
they form the three families of SM matter fields together.
The gauge invariance is reduced to the SM one at the origin on $\Omega$,
because the gauge transformation function compatible with the above condition (\ref{conf})
is constrained such that
\begin{eqnarray}
&~& \xi^{a}(x,0) \ne 0 , ~~ \xi^{\hat{a}}(x,0) = 0 .
\label{xi}
\end{eqnarray}
Note that $SU(5)$ gauge symmetry exists where $\xi^{a}(x,w^2) \ne 0$ and $\xi^{\hat{a}}(x,w^2) \ne 0$.
The essence of a reduction to the SM is that extra fields vanish at the origin on $\Omega$.
It is not trivial whether or not a field, which is an even function of $y^j$,  vanishes  at $y^j = 0$.
Furthermore, it is necessary to stabilize field configurations around the origin at the quantum level.

\subsection{Emergent MSSM}

The SM has been established as an effective theory on a weak scale, but it suffers from several problems.
Typical ones are the naturalness problem (How is the weak scale stabilized?) 
and the dark matter problem (What is dark matter?).
These problems are solved by the introduction of space-time supersymmetry.
As a bonus, gauge coupling unification occurs on the basis of the MSSM
with a large desert hypothesis between the TeV scale and the grand unification scale.\cite{GCU}

Let us discuss the space-time SUSY extension of our model. 
Our main players are BRST superfields of the vector superfield $V=(A_M; \lambda_1, \lambda_2)$, 
Higgs hyperfields $(H_1, H^c_1) = (h_1, \tilde{h}_1, h_1^c, \tilde{h}_1^c)$, 
$(H_2, H^c_2) = (h_2, \tilde{h}_2, h_2^c, \tilde{h}_2^c)$ 
and $n_g$ families of matter hyperfields 
$\Psi_{\bm n} = (\tilde{\psi}_{\bm n}, \tilde{\psi}_{\bm n}^c; \psi_{\bm n}, \psi_{\bm n}^c)$.
Here, $\lambda_1$ and $\lambda_2$ are gauginos, $\tilde{h}_1^{(c)}$ and $\tilde{h}_2^{(c)}$ are higginos,
and the $\tilde{\psi}_{\bm n}$ are sfermions (superpartners of matter fields).
As global symmetries, there are $SU(2)_R$ symmetry 
and flavor symmetry rotating Higgs hypermultiplets ($SU(2)_H$).

We assume that the space-time SUSY is partially broken by some dynamics and
the following action appears at $M_U$:
\begin{eqnarray}
\hspace{-1cm}&~& S_{SU(5)}^{\tiny{\rm SUSY}} = \int d^4x \int d^2y \int d\theta d\bar{\theta} {\mathcal{L}}(x, w^2) ,
\nonumber \\
\hspace{-1cm}&~&   {\mathcal{L}}(x, w^2)  =  -\frac{1}{2}{\rm tr}F^{\mu\nu}(x,w^2)F_{\mu\nu}(x,w^2)
 + \sum_{s=1,2} {\rm tr}\left(i\bar{\lambda}_s(x, w^2) \Gamma^{\mu} D_{\mu} \lambda_s(x, w^2)\right)
\nonumber \\
\hspace{-1cm}&~& ~~~ + \sum_{s=1,2} \left(|D_{\mu} h_s(x,w^2)|^2 + |D_{\mu} h_s^c(x,w^2)|^2\right)
\nonumber \\
\hspace{-1cm}&~& ~~~ + \sum_{s=1,2} \left(i\bar{\tilde{h}}_s(x,w^2) \Gamma^{\mu} D_{\mu} \tilde{h}_s(x,w^2)
 + i\bar{\tilde{h}}_s^c(x,w^2) \Gamma^{\mu} D_{\mu} \tilde{h}_s^c(x,w^2)\right)
 + \cdots
\nonumber \\
\hspace{-1cm}&~& ~~~ + \sum_{n_g~{\rm families}} 
(i \bar{\psi}_{\bar{\bf 5}}(x,w^2) \Gamma^{\mu} D_{\mu} \psi_{\bar{\bf 5}}(x,w^2)
 + i \bar{\psi}_{{\bf 10}}(x,w^2) \Gamma^{\mu} D_{\mu} \psi_{{\bf 10}}(x,w^2)
\nonumber \\
\hspace{-1cm}&~& ~~~ + i \bar{\psi}^c_{\bar{\bf 5}}(x,w^2) \Gamma^{\mu} D_{\mu} \psi^c_{\bar{\bf 5}}(x,w^2)
 + i \bar{\psi}^c_{{\bf 10}}(x,w^2) \Gamma^{\mu} D_{\mu} \psi^c_{{\bf 10}}(x,w^2)
\nonumber \\
\hspace{-1cm}&~& ~~~ + |D_{\mu} \tilde{\psi}_{\bar{\bf 5}}(x,w^2)|^2 + |D_{\mu} \tilde{\psi}_{\bf 10}(x,w^2)|^2
 + |D_{\mu} \tilde{\psi}^c_{\bar{\bf 5}}(x,w^2)|^2 + |D_{\mu} \tilde{\psi}^c_{\bf 10}(x,w^2)|^2
\nonumber \\
\hspace{-1cm}&~& ~~~ + f_{U} h(x,w^2) {\psi}_{{\bf 10}}(x,w^2) {\psi}_{{\bf 10}}(x,w^2) 
+ f_{D} h^{*}(x,w^2) {\psi}_{{\bf 10}}(x,w^2) \psi_{\bar{\bf 5}}(x,w^2) + {\rm h.c.} 
\nonumber \\
\hspace{-1cm}&~& ~~~ + \cdots) .
\label{S-8D-SUSY}
\end{eqnarray}
The above action (\ref{S-8D-SUSY}) is still $OSp(2/2)$-invariant and possesses the space-time SUSY
(corresponding $N=1$ SUSY in 4D).
The MSSM fields are $A^a_{\mu}$, $\lambda^a_1$, $h_{W1}$, $h_{W2}$, $\tilde{h}_{W1}$, $\tilde{h}_{W2}$
and three families of matter chiral multiplets.
If ${\mathcal{L}}(x, w^2)$ vanishes at infinity on $\Omega$ and only the MSSM particles survive at the origin on $\Omega$,
the MSSM action without the $\mu$ term can be derived after dimensional reduction.
The $\mu$ term is forbidden by the $U(1)_R$ symmetry.
The $U(1)_R$ symmetry is the diagonal subgroup of $U(1)$ symmetries in $SU(2)_R$ and $SU(2)_H$,
and the $U(1)_R$ charge ($Q_R$) is assigned as $Q_R(V) = Q_R(H_s) = 0$, $Q_R(\Psi_{\bm n}) = 1$
and $Q_R(H_s^c) = 2$.\cite{HN}
The $\mu$ term with a suitable magnitude could be generated on the breakdown of space-time SUSY, e.g.,
through the Guidice-Masiero mechanism.\cite{GM} 
The $U(1)_R$ symmetry also forbids dangerous proton decay processes 
through operators of dimensionality four.

\section{Conclusions and discussion}

We have proposed a new grand unification scenario for ensuring 
proton stability and triplet-doublet Higgs mass splitting 
with the help of topological symmetry and dynamics.
We have given simple models with an $SU(5)$ gauge group on an extended space-time with extra superspace
to realize our proposal.
Dangerous colored particles are eliminated
and the reduction to the SM or MSSM is carried out using the Parisi-Sourlas mechanism
under nontrivial ansatz for field configurations.

Since extra particles are unphysical in topological grand unification, 
it could be checked using the MSSM particles's features and their effects.
Gaugino and sfermion masses can be useful probes
because the unification of particles occurs 
and those masses should agree at $M_U$ 
among members in each $G_U$-multiplet.\cite{KMY}

There are open questions for our models and/or scenario.
Is it possible to arrange field configurations in order to realize the SM or MSSM?
If possible, how are they fixed and are they stable against radiative corrections?
Our starting models could be reformulated as gauge theories with stochastic classical dynamics.
The key solving riddles might be hidden there.
Is it possible to realize unphysical grand unification in a more fundamental theory such as
the superstring theory?
Even if our models do not work for any reason, 
study of elementary particle physics using a topological symmetry\footnote{
An $OSp(4/2)$ supersymmetric model and a deformation of the topological field theory
were studied to explore the quark confinement mechanism in QCD.\cite{HK,Kondo}}
would be attractive and it is worth exploring a high-energy theory on the basis of our proposal.

\section*{Acknowledgements}
This work was supported in part by scientific grants from the Ministry of Education, Culture,
Sports, Science and Technology under 
Grant Nos.~18204024 and 18540259.

\end{document}